\documentclass[a4paper]{article}

\usepackage{caption} 
\usepackage{tabularx}
\usepackage{float}
\usepackage{array}  
\usepackage{booktabs} 
\usepackage{multirow}
\usepackage{makecell}
\usepackage{arydshln} 
\usepackage{color}
\usepackage{bbding}
\usepackage{amsfonts,amssymb} 
\usepackage{subfigure} 
\usepackage{bbding}  

\usepackage{INTERSPEECH2022}

\title{Event-related data conditioning for acoustic event classification}
\name{Yuanbo Hou, Dick Botteldooren}
\address{WAVES, Ghent University, Belgium}
\email{Yuanbo.Hou@UGent.be, Dick.Botteldooren@UGent.be}

\begin{document}

\maketitle 
\begin{abstract}
Models based on diverse attention mechanisms have recently shined in tasks related to acoustic event classification (AEC). Among them, self-attention is often used in audio-only tasks to help the model recognize different acoustic events. Self-attention relies on the similarity between time frames, and uses global information from the whole segment to highlight specific features within a frame. 
In real life,  information related to acoustic events will attenuate over time, which means the information within some frames around the event deserves more attention than distant time global information that may be unrelated to the event. 
This paper shows that self-attention may over-enhance certain segments of audio representations, and smooth out the boundaries between events representations and background noises. Hence, this paper proposes an event-related data conditioning (EDC) for AEC. EDC directly works on spectrograms. 
The idea of EDC is to adaptively select the frame-related attention range based on acoustic features, and gather the event-related local information to represent the frame.
Experiments show that: 1) compared with spectrogram-based data augmentation methods and trainable feature weighting and self-attention, EDC outperforms them in both the original-size mode and the augmented mode; 2) EDC effectively gathers event-related local information and enhances boundaries between events and backgrounds, improving the performance of AEC.

\end{abstract}
\noindent\textbf{Index Terms}: Acoustic event classification, event-related data conditioning, SpecAugment, mixup, feature weighting

\section{Introduction}

Acoustic event classification (AEC) performs multi-label classification on audio clips, which aims to identify whether target acoustic events occur in the audio clip.
AEC can help the agent to understand its surrounding environment and what is happening. 
Classical applications of AEC include security surveillance \cite{medical_surveillance},  robot hearing \cite{ren2016sound}, monitoring \cite{monitor} and smart cities \cite{smart, almaadeed2018automatic}.

Previous methods for AEC are often based on multilayer perceptron architecture \cite{mlp}, convolutional neural networks (CNN) \cite{cnn, aec_Phan}, recurrent neural networks (RNN) \cite{RNN}, and convolutional recurrent neural networks (CRNN) \cite{crnn}. 
Recently, motivated by the excellent performance of the attention-based Transformer \cite{transformer} on sequence-related tasks, various attention mechanisms \cite{att1, att2, att3}, especially self-attention, have been widely used in audio-related tasks \cite{ast}.
Self-attention \cite{transformer} shows a good balance between the ability to model long-term dependencies and computational efficiency \cite{long_term}.
Memory controlled self-attention is used to model relations between tokens within sound events to improve sound recognition \cite{memory}. And models with self-attention can be combined with connectionist temporal classification (CTC) loss to enforce a monotonic ordering and utilize timing information of acoustic events in audio clips \cite{allinone}.
CNN-Transformer is also proposed for audio event tasks, which performs similarly to the CRNN, but has the advantage of being efficiently trained in parallel \cite{cnn_t}.

In audio-related Transformer-based models \cite{ast, audio_trans, Mei2021b, audio_trans2, audio_trans3}, self-attention measures the cross-frame similarity within the whole segment and obtains new representations by weighted summation.
In the above-mentioned models related to acoustic events, during the training, self-attention blindly transforms the frame based on global information from the whole segment, regardless of whether the information is related or unrelated to events in the frame.  
The globally calculated self-attention, also known as scaled dot-product attention \cite{transformer}, has the following drawbacks.
1) The information attenuation of sound events over time in real life is not considered.
In real life, audio events  usually last a few seconds. For some daily  events, the shortest duration is about $0.5s$ \cite{Hou2017}.  
And events containing fluctuating or intermittent sounds (e.g., dogs barking, birds singing) usually consist of short events.
That is, sound events in real life usually last a few seconds, and they are more related to other events in the neighboring time than events in a distant time. So the local information around the event deserves more attention.
2) After weighting and representing the frame based on global information, other events' information that may be irrelevant to acoustic events in the frame will be added. That is, self-attention may not be able to effectively focus on the event-related local information, and cannot prevent the introduction of irrelevant information into representations of acoustic events. 
3) Reconstructing local features based on global attention will make the local features approach global representations, resulting in boundaries between acoustic events and backgrounds being smoothed out in audio clips. It is difficult for the model to judge where the information of events is located, which increases the learning burden of the model.
These shortcomings of self-attention are inconsistent with the characteristics of audio events.
To mitigate the three deficiencies mentioned above, this paper proposes an event-related data conditioning (EDC) for AEC. 
EDC automatically selects the event-related attention range for each frame, locally focusing on related information without introducing features of unrelated distant time sounds into the representation.

The proposed data conditioning method EDC directly acts on spectrograms and does not require training, which is similar to spectrogram-based augmentation methods: SpecAugment \cite{spec} and Mixup \cite{mixup}.
SpecAugment introduces artificial noises on spectrograms to diversify training samples. Mixup randomly mixes spectrograms of two different samples in proportion to form a new sample.
These data augmentation methods make the model focus on core features and ignore irrelevant details by creating diverse training samples with the same relevant features.
The EDC will be compared with SpecAugment and Mixup. 
The essence of EDC is deterministic event-related feature transformation.
And feature weighting \cite{feature_w}, which is dedicated to dynamically adapting learnable features, also performs well in the AEC task. Therefore, this paper will also compare and demonstrate the performance of feature weighting.
Even though EDC, Mixup, SpecAugment, and feature weighting are different approaches, their underlying goals are the same: to make the model less sensitive to irrelevant features of target events and attempt to preserve key features.
The proposed EDC is inspired by the performance analysis of self-attention in AEC tasks, so EDC that does not require training will be compared with the trainable self-attention to show the difference.

This paper proposes EDC, which is a deterministic feature transformation to gather event-related local information and enhance boundaries between events and backgrounds.
This paper is organized as follows, Section 2 introduces the proposed EDC. Section 3 describes the dataset, model, experimental setup, and analyzes the results. Section 4 gives conclusions.

\section{Proposed method}
\label{sec:format}

The acoustic event classification task, and the proposed event-related data conditioning method, are introduced in this part.

\subsection{Acoustic event classification (AEC)}
AEC performs multi-label classification on audio clips to identify and recognize target events \cite{aec_mark}. In acoustic event-related tasks, the most frequently used efficient acoustic feature is the log mel spectrogram  \cite{mlp, RNN, cnn_t, mixup, Hou2018}.
The input audio clip $x$ is transformed to the time-frequency representation $X(t; f)$ of log mel spectrogram. 
The purpose of AEC is to train a model to play the role of classification mapping: $X(t; f)\Rightarrow Y$, where $Y=[y_1,..,y_K], K$ is the number of classes of audio events, and $y_k \in [0,1]$ is the presence probability of the $k$-th event in the audio clip, $k=1,..,K$.

\subsection{Event-related data conditioning (EDC)}

\begin{figure}[t]
	\setlength{\belowcaptionskip}{-0.3cm}   
	\centerline{\includegraphics[width = 0.5\textwidth]{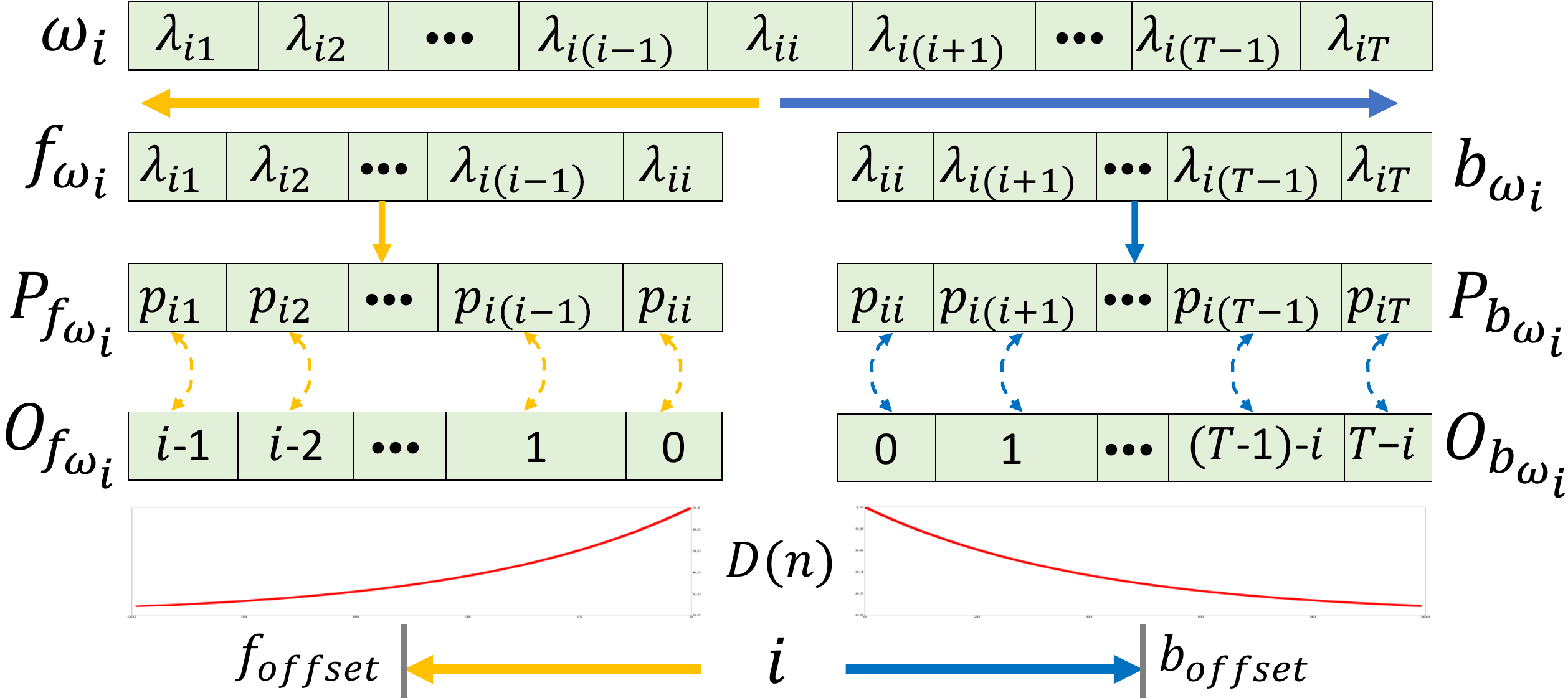}}
	\caption{The calculation procedure of the proposed EDC.}
	\label{EDC}
\end{figure}

The process of EDC is shown in Figure~\ref{EDC}.
Let $\boldsymbol{X}\in \mathbb{R}^{T \times F}$ denotes the acoustic time-frequency representation of the input audio clip, where $\boldsymbol{x_i}\in\mathbb{R}^{1 \times F}$ is the $i$-th frame, $T$ and $F$ denote the number of time frames and the acoustic feature dimension, respectively. 
For $\boldsymbol{X}$, the corresponding \textit{cross-frame} \textit{similarity} \textit{matrix} \cite{guo2021ssan} $\Omega$ is obtained by: 
\begin{equation}
\setlength{\abovedisplayskip}{3pt}
\setlength{\belowdisplayskip}{3pt} 
\Omega = \boldsymbol{X}\boldsymbol{X}^T = 
\begin{bmatrix}
	      \boldsymbol{x_1}\boldsymbol{x_1}
	      &...&\boldsymbol{x_1}\boldsymbol{x_T}\\
	      \boldsymbol{x_2}\boldsymbol{x_1}
	      &...&\boldsymbol{x_2}\boldsymbol{x_T}\\
	      ... 
	      &  ... & ... \\
	      \boldsymbol{x_T}\boldsymbol{x_1} 
	      &...&\boldsymbol{x_T}\boldsymbol{x_T}\\
	      \end{bmatrix}              
  = \begin{bmatrix}
	       \omega_1\\
	      \omega_2\\
	      ...\\
	      \omega_T\\
	      \end{bmatrix} 
\end{equation}
\begin{equation}
\setlength{\abovedisplayskip}{3pt}
\setlength{\belowdisplayskip}{3pt}  
\omega_i = [\boldsymbol{x_i}\boldsymbol{x_1}, \boldsymbol{x_i}\boldsymbol{x_2}, ..., \boldsymbol{x_i}\boldsymbol{x_T}] = [\lambda_{i1}, ..., \lambda_{iT}] 
\end{equation}
where the row vector $\{\omega_i\in\mathbb{R}^{T}, i\in[1,T]\}$ is the similarity vector of the $i$-th frame. 
Specifically, $\{\lambda_{ij}=\boldsymbol{x_i}\boldsymbol{x_j}, i\in[1,T], j\in[1,T]\}$ denotes the similarity value between the $i$-th frame and the $j$-th frame. Next, for the similarity vector $\omega_i$ of the $i$-th frame, the following procedure is proposed:

\noindent\textit{\textbf{1) Divide the $\omega_i$ in a forward and a backward similarity vector}}

To consider the similarity between the current frame and previous frames and following frames respectively, the similarity vector $\omega_i$ of the  $i$-th frame in the $\Omega$ is divided into the \textit{forward} similarity vector $f_{\omega i}$,
\begin{equation}
\setlength{\abovedisplayskip}{3pt}
\setlength{\belowdisplayskip}{3pt}
f_{\omega i} = [\lambda_{i1}, \lambda_{i2}, ..., \lambda_{i(i-1)}, \lambda_{ii}] 
\end{equation}
and the \textit{backward} similarity vector $b_{\omega i}$.
\begin{equation}
\setlength{\abovedisplayskip}{3pt}
\setlength{\belowdisplayskip}{3pt}
b_{\omega i} = [\lambda_{ii}, \lambda_{i(i+1)},..., \lambda_{i(T-1)}, \lambda_{iT}]
\end{equation}
Next, taking the \textit{forward} similarity vector $f_{\omega i}$ as an example.
\\\textit{\textbf{2) Estimate the time range of the frame-related similar information}}

The \textit{row-wise} softmax \cite{transformer} is applied to $f_{\omega i}$ to obtain the corresponding similarity probability distribution $P_{f_{\omega i}}$,
\begin{equation}
\setlength{\abovedisplayskip}{3pt}
\setlength{\belowdisplayskip}{3pt}
P_{f_{\omega i}} = softmax(f_{\omega i}) = [p_{i1}, p_{i2}, ..., p_{ii}]
\end{equation}
then $P_{f_{\omega i}}$ is viewed as the probability of the frame-related similarity offset value distribution. The offset value of each previous frame position relative to the  $i$-th frame and
corresponding to the $f_{\omega i}$ is $O_{f_{\omega i}} = [i-1, i-2, ..., 1, 0]$. Hence, based on the distribution probability of similarity at offset values, the expectation value of the offset is 
\begin{equation}
\setlength{\abovedisplayskip}{3pt}
\setlength{\belowdisplayskip}{3pt}
f_{offset} = E(O_{f_{\omega i}}) = \sum\nolimits_{n=1}^{i} O_{f_{\omega i}}(n) P_{f_{\omega i}}(n)
\end{equation}
The similarity value in $[i-f_{offset}, i]$ on the time axis is regarded as effective similar information related to the $i$-th frame. This definition of $f_{offset}$ results in unwanted effects in the case of multiple similar events that cause the probability distribution to show multiple peaks. Therefore a further step is introduced.  
\\\textit{\textbf{3) Weight the effective similarity time range}}

\label{ssec:EDC}

As time increases, the probability that similar features belong to the same event decreases. Thus, independently of the probability of similarity $P_{f_{\omega i}}$, the weighting of offsets should decrease. This knowledge is simply introduced by \\
\begin{equation}\label{factor}
\setlength{\abovedisplayskip}{2pt}
\setlength{\belowdisplayskip}{3pt}
f_{offset} = \sum\nolimits_{n=1}^{i} O_{f_{\omega i}}(n) P_{f_{\omega i}}(n) D(n)
\end{equation}
where $D(n)=e^{-\alpha (i-n)}$ is the attenuation coefficient over time, $\alpha$ is the attenuation factor. 
The smaller $\alpha$, the smaller the selectable time range.  
The effect of $\alpha$ on the AEC task will be analyzed in detail later in the experiment.

Based on the way of calculating the \textit{forward} effective time range, the \textit{backward} effective time range $[i, i+b_{offset}]$ is also obtained, where $b_{offset}$ is the \textit{backward} corrected expectation value of offset. For the $i$-th frame, the final effective similarity time range is $[i-f_{offset}, i+b_{offset}]$. 

After estimating the effective time range of all frames, an \textit{effective} \textit{similarity} \textit{time} range matrix $\Phi$ is obtained. $\Phi$ is composed of $-\infty$ and 1, where similarity values outside the effective time range will be masked with $-\infty$ following the settings in \cite{transformer}. The \textit{cross-frame} \textit{similarity} \textit{matrix} $\Omega$ is truncated based on $\Phi$ and normalized by \textit{row-wise} softmax,
\begin{equation}
\setlength{\abovedisplayskip}{3pt}
\setlength{\belowdisplayskip}{3pt}
EDC_{output} =softmax(\Omega \odot \Phi) \boldsymbol{X}
\end{equation}
where the symbol $\odot$ is the element-wise product. The EDC acts directly on the spectrogram $X$, similar to SpecAugment and Mixup, which enhance data based on spectrograms. 
Later experiments compare the performance of EDC, SpecAugment, and Mixup in original-size mode (\textit{OM}) and augmented mode (\textit{AM}). 
Denote augmented features after EDC, SpecAugment, and Mixup as $\hat{X}_E$, $\hat{X}_S$, $\hat{X}_M$, respectively.
In the \textit{OM}, the model is trained on $\{X, \hat{X}_E, \hat{X}_S, \hat{X}_M\}$, which is consistent with the amount of original training data, to compare differences between original features and enhanced features.
In the \textit{AM}, the model is trained on $\{(X, X), (X, \hat{X}_E), (X, \hat{X}_S), (X, \hat{X}_M)\}$,
similar to the data augmentation method, which is equivalent to doubling the amount of training data.

\section{Experiments and results}
\vspace{-0.1cm}
\subsection{Dataset, Model, and Experiments Setup}

To evaluate the performance of the proposed EDC, an acoustic event-related deterministic feature transformation method, the large-scale weakly labeled DCASE domestic environment acoustic event dataset \cite{DCASe2018} and the CHiME-Home \cite{chime} dataset for investigating the in-home sound source recognition are used in this paper for the AEC task. 
The audio clip in the DCASE dataset is 10-second, and the audio clip in the CHiME-Home is 4-second. 
The DCASE domestic audio dataset excerpted from Audioset \cite{aduioset} contains 10 classes of real-life polyphonic audio events. And 7 classes of different in-home audio events are annotated in the CHiME-Home dataset.

\begin{figure}[t] 
	\setlength{\abovecaptionskip}{0.cm}    
	\setlength{\belowcaptionskip}{-0.5cm}   
	\centerline{\includegraphics[width = 0.215\textwidth, angle=90]{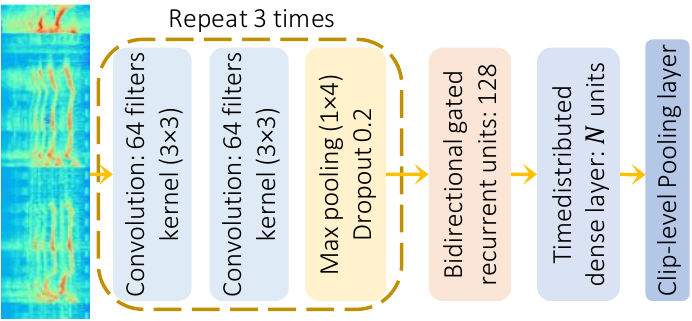}} 
\caption{Structure of the cornerstone model used in this paper.}
	\label{model}
\end{figure}

Convolutional recurrent neural networks (CRNN) perform well in AEC-related tasks \cite{crnn, Hou2018, RNN}, hence a typical CRNN in Figure~\ref{model} is used as the cornerstone model in this paper to test the performance of different data conditioning methods.
In Figure~\ref{model}, convolutional layers are applied to learn local shift-invariant patterns from features. 
To preserve the time resolution of the input, pooling is applied to the frequency axis only. 
Bidirectional gated recurrent units (BGRU) \cite{RNN} are adopted to capture the temporal context information. 
The timedistributed dense layer has $N$ units, where \textsl{N} is the number of acoustic event classes. 
When using the DCASE and CHiME datasets, $N$ equals 10 and 7, respectively.
In the last pooling layer, global max pooling (\textit{GMP}) or global average pooling (\textit{GAP}) is optional. 
The previous study \cite{Hou2018} shows that \textit{GMP} performs better than \textit{GAP} in the AEC task. 
The reason may be that \textit{GMP} encourages the response for a single location to be high, while \textit{GAP} encourages all responses  to be high \cite{gmap}. 
An intuitive explanation based on the observation is the audio event does not appear in every frame in audio clips. 
As long as the event exists in certain frames, \textit{GMP} can locate it and thus is more suitable for the AEC task. 
Therefore, \textit{GMP} is used in the last pooling layer, from which clip-level probabilities of acoustic events can be obtained. 
Since EDC is inspired by self-attention, the trainable self-attention will be added after the convolutional layer of the model to compare with EDC. 
For more details and the source code, please see the project homepage\protect\footnote{https://github.com/Yuanbo2020/EDC}.

\label{ssec:model}

In training, log mel-bank energy with 64 banks \cite{mel} is used as input acoustic features, which is extracted by STFT with Hamming window length of 40\textit{ms} and overlap of 50\% between the window. 
Dropout and normalization are used to prevent over-fitting. 
The loss function is binary cross-entropy, Adam optimizer \cite{adam} with a learning rate of 0.001 is used to minimize the loss.
Under the same training conditions, the amount of training data for \textit{AM} is twice that of \textit{OM}. 
Therefore, to avoid the interference of different data amounts for model performance evaluation, early stopping is adopted in training. Specifically, after 50 epochs of training, if the performance of the model on the validation set does not improve within 20 epochs, the training will be stopped.
All systems are trained on a single graphical card (Tesla V100-SXM2-32GB) with a batch size of 64 for maximum 500 epochs.
To avoid the influence of different threshold interference, the area under curve (\textsl{AUC}) \cite{auc}  without threshold is used as the metric to comprehensively measure the results.
Bigger \textit{AUC} indicates better performance.

\begin{table}[b]\footnotesize
	\setlength{\abovecaptionskip}{0cm}   
	\setlength{\belowcaptionskip}{-0.45cm}   
	\renewcommand\tabcolsep{2.0pt} 
	\centering
	\caption{Maximum time range (\textit{frames}) and \textit{AUC} at different $\alpha$.}
	\begin{tabular}
	{
	p{1cm}<{\centering} |
	p{1.2cm}<{\centering} |
	p{1.2cm}<{\centering}
	p{1.2cm}<{\centering}|
	p{1.2cm}<{\centering}
	p{1.2cm}<{\centering}
	} 
		   
\hline
		
		\multirow{2}{*}{\textit{$\alpha$}} 
		& \multirow{2}{*}{\textit{frames}}  
		& \multicolumn{2}{c|}{\textit{AUC (\%) of OM}}
		& \multicolumn{2}{c}{\textit{AUC (\%) of AM}}  
		\\
		
		\cline{3-6}
		
		  &   & \textit{DCASE} & \textit{CHiME} & \textit{DCASE} & \textit{CHiME} \\
		
		\cline{1-6} 
		2.5 & 18 & 93.68 & 95.10 & 94.23 & 96.80 \\
		
		5 & 38 & 94.01 & 96.36 & 95.36 & 96.81 \\
		
		7 & 54 & \textit{\textbf{95.20}} & 96.71 & \textit{\textbf{96.03}}  & 97.56 \\
		
		10 & 78 & 93.93 & \textit{\textbf{96.93}} & 95.13 & \textit{\textbf{98.10}} \\
		
		20 & 156 & 93.12 & 95.71 & 93.62 & 97.45 \\
		
		50 & 390 & 90.79 & 92.98 & 92.13 & 97.38 \\
		
		100 & 782 & 89.83 & 91.69 & 91.88 & 96.74 \\
		
		250 & 1956 & 88.11 & 89.16 & 88.73 & 95.86 \\
		
		500 & 3912 & 82.01 & 88.39 & 84.48 & 95.64  \\
		

\hline
		
	\end{tabular}
	\label{tab:rule}
\end{table}

\label{ssec:tsne}
\begin{figure}[t]
	\setlength{\abovecaptionskip}{0.1cm}   
	\setlength{\belowcaptionskip}{-0.6cm}   
	\centerline{\includegraphics[width = 0.5\textwidth]{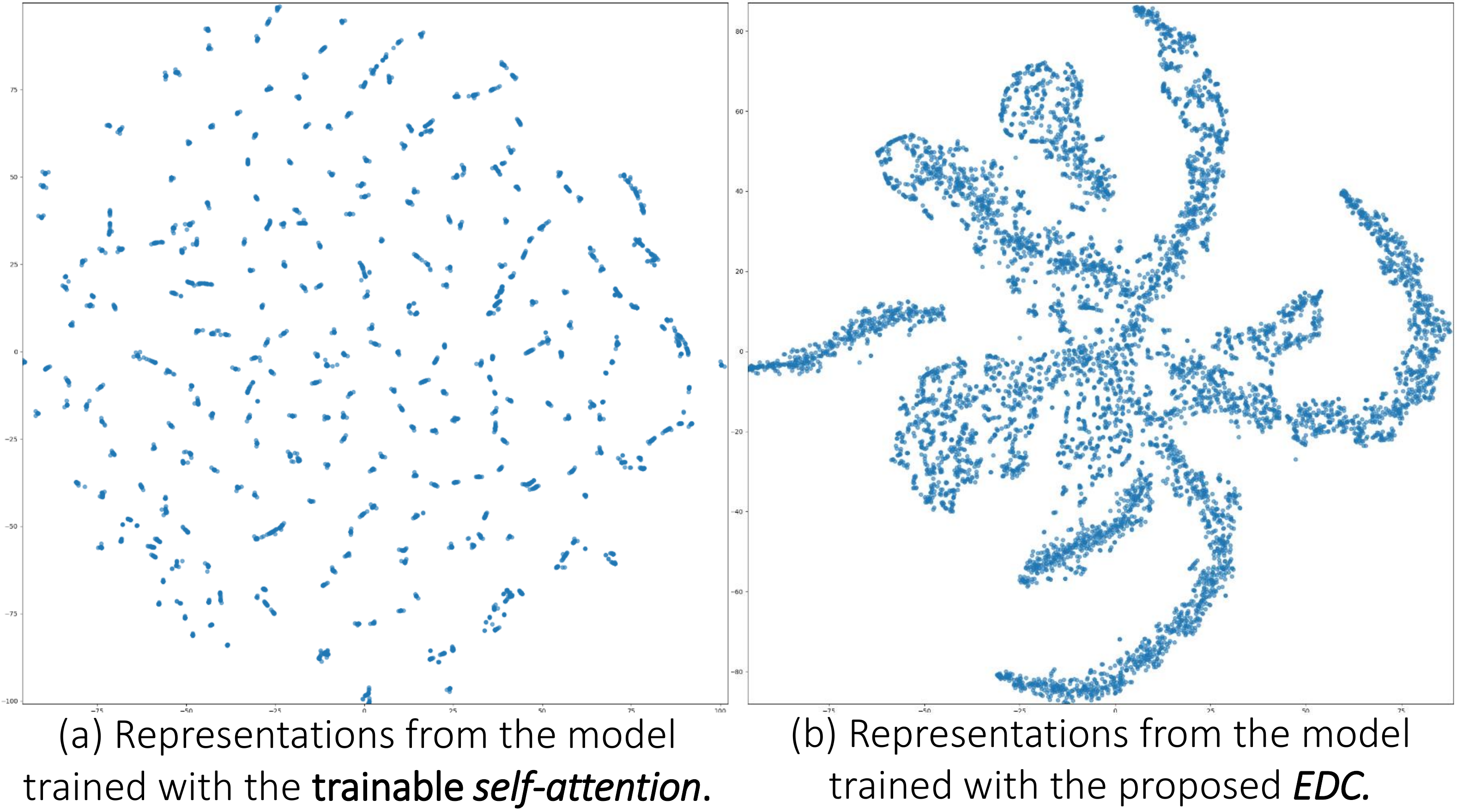}}
	\caption{Visualization of frame-level representations distribution using unsupervised \textit{t-SNE} \cite{tsne}. Please note that models in this paper are trained by clip-level weak labels in datasets of \textit{DCASE} and \textit{CHiME}, and the label of each audio clip is a multi-hot vector, so the label corresponding to the frame-level representation is unknowable.}
	\label{tsne}
\end{figure}

\vspace{-0.15cm}
\subsection{Results and Analysis}
As shown in Equation~\ref{factor}, the attenuation factor $\alpha$ affects the selectable time range. So the first \textbf{r}esearch \textbf{q}uestion (RQ) is:
\\\textbf{• RQ1}: What impact will different $\alpha$ values have?

In experiments, raw audio clips are converted into frames with a hop size of $20ms$ between frames, resulting in the number of frames of input acoustic features of datasets DCASE and CHiME being 500 and 200, respectively.
As shown in Table \ref{tab:rule}, $\alpha$ at $7$ and $10$ achieved best AEC results, the corresponding maximum selectable time range is 54 frames ($1.08s$) and 78 frames ($1.56s$). 
That is, for acoustic events in the two datasets, paying attention to information within $0.54s\sim0.78s$ around their time position is the most effective.  
Previous statistics  \cite{Hou2017} show the shortest duration of some daily audio events in real life is about $0.5s$ $(0.29s\sim0.53s)$. 
And events containing fluctuating or intermittent sounds usually consist of short events. 
Therefore, if the model can identify short-term acoustic events, it is able to recognize these events in a long term.

When $\alpha$ is 500, the information within 1956 frames before and after the frame will be considered in its scope of attention. The number of frames of samples in the DCASE dataset is 500, so when $\alpha$ is 500, the effect of attenuation coefficient is weakened, and EDC tends to focus on global information instead of local parts, the EDC cannot effectively focus on the event-related local information, so its performance is poor.

\noindent
\textbf{• RQ2}: Compared with feature weighting, SpecAugment, Mixup and the trainable self-attention, how does EDC perform?

SpecAugment has three specific ways \cite{spec}: time warping, frequency masking, and time masking. 
The ablation study in Table \ref{tab:ablation} shows under the same training condition, the performance of introducing a single type of noise is similar. If more types of noise are introduced, the performance of the model is worse. The reason may be that more types of noise will increase the learning burden of the model, extracting knowledge from noisy data at the same time is challenging.

\begin{table}[b]\footnotesize
	\setlength{\abovecaptionskip}{0cm}   
	\setlength{\belowcaptionskip}{-0.4cm}   
	\renewcommand\tabcolsep{2.0pt} 
	\centering
	\caption{Ablation study of SpecAugment for the AEC task.}
	\begin{tabular}
	{
	p{0.8cm}<{\centering} |
	p{1.2cm}<{\centering} |
	p{0.8cm}<{\centering} |
	p{1.05cm}<{\centering}
	p{1.05cm}<{\centering}|
	p{1.05cm}<{\centering}
	p{1.05cm}<{\centering}
	} 
\hline
		\textit{Time} 
		& \textit{Frequency}  
		& \textit{Time}  
		& \multicolumn{2}{c|}{\textit{AUC (\%) of OM}}
		& \multicolumn{2}{c}{\textit{AUC (\%) of AM}}  
		\\
		
		\cline{4-7}
		
		 \textit{mask} & \textit{mask} & \textit{warp}& \textit{DCASE} & \textit{CHiME} & \textit{DCASE} & \textit{CHiME} \\
		
		\cline{1-7} 
		\specialrule{0pt}{1pt}{0pt}
		\XSolidBrush & \XSolidBrush & \XSolidBrush & 90.55 & 91.28 & 90.71 & 91.42\\
		
		\Checkmark & \XSolidBrush & \XSolidBrush & 91.41 & 95.04 & 94.27 & \textbf{\textit{96.62}}\\
		
		 \XSolidBrush & \Checkmark & \XSolidBrush & 92.96 & \textbf{\textit{96.24}} & \textbf{\textit{94.37}} & 96.10\\
		
		\XSolidBrush& \XSolidBrush & \Checkmark & \textbf{\textit{94.46}} & 95.50 & 94.13 & 96.25\\
		
		\Checkmark& \Checkmark & \XSolidBrush& 92.58 & 94.58 & 93.74 & 95.28\\
		
		\Checkmark & \XSolidBrush & \Checkmark & 91.59 & 93.72 & 93.07 & 95.52\\
		
		\XSolidBrush& \Checkmark & \Checkmark & 92.18 & 94.43 & 93.80 & 95.75\\
		
		\Checkmark & \Checkmark & \Checkmark & 92.61 & 94.83 & 93.06 & 94.94\\
\hline
		
	\end{tabular}
	\label{tab:ablation}
\end{table}

\label{ssec:figure-f}
\begin{figure}[t] 
	\setlength{\abovecaptionskip}{0.05cm}    
	\setlength{\belowcaptionskip}{-0.5cm}   
	\centerline{\includegraphics[width = 0.5  \textwidth]{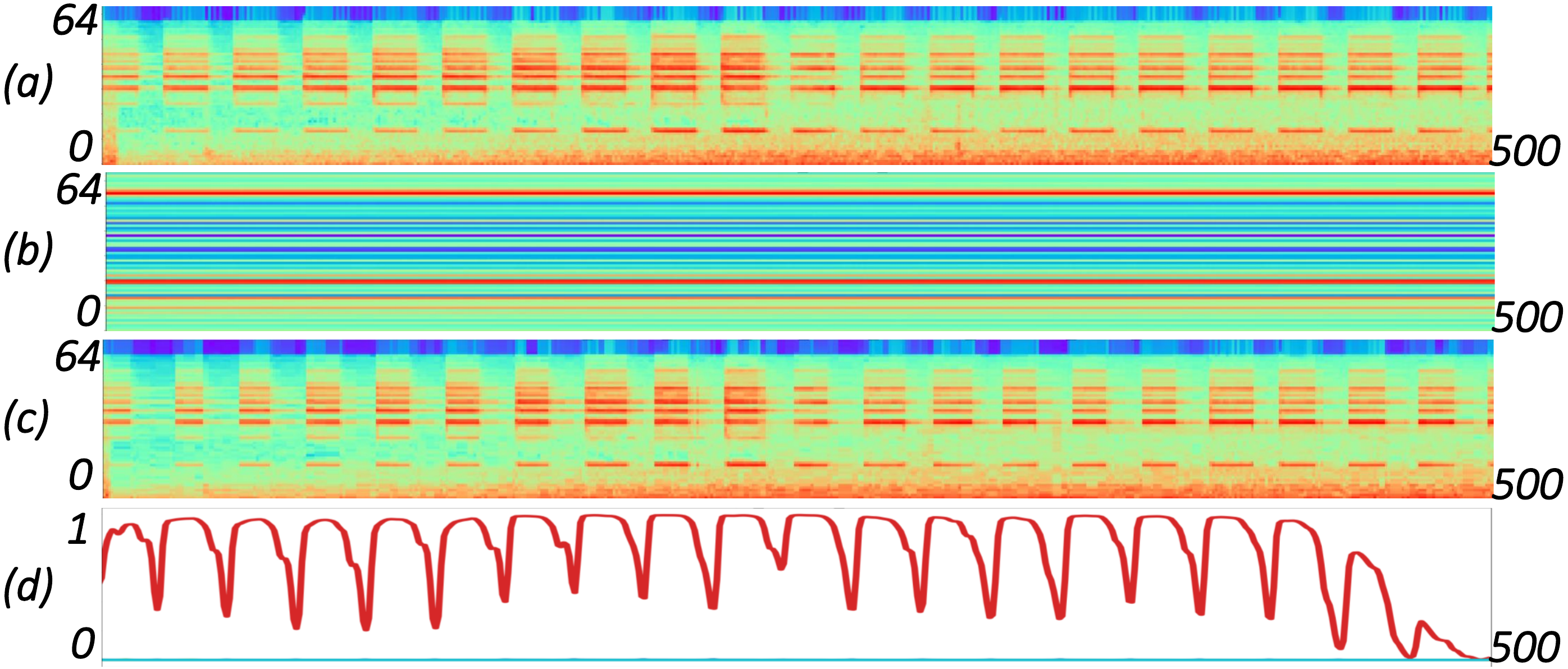}}
	\caption{A demo of the siren sound clip. Subgraph: (a) Log mel spectrogram; (b) Bottleneck features from the trainable self-attention layer; (c) Acoustic features after EDC; (d) The probability of events predicted by the model trained with EDC.  
	}
	\label{figure-f}
\end{figure}

In Table \ref{tab:models}, Mixup performs better than SpecAugment. 
The reason may be that SpecAugment diversifies the training data, but Mixup diversifies training data and labels.
Unlike EDC and SpecAugment, Mixup modifies labels while modifying spectrograms.
The model with trainable self-attention performs worse than that with EDC.
Table \ref{tab:rule} also shows the trend that results of EDC becomes worse as the attention selectable time range for each frame becomes wider. In addition,  Figure~\ref{tsne} shows that compared with the loosely distributed samples from the model with self-attention, the representations of the model trained by EDC can be automatically clustered into different clusters, which can greatly reduce the classification burden of the model.
The overall performance of feature weighting (\# 5 in Table \ref{tab:models}) is better than that of methods other than the proposed EDC. 
The reason may be that feature weighting can dynamically weight global features in different scales based on their importance, and aggregate cross-scale features for the AEC task \cite{feature_w}.

Overall, the proposed EDC, which focuses on gathering event-related local information and enhancing boundaries between events and backgrounds, performs best in Table \ref{tab:models}.
The reason may be that the balance of EDC between disregarding the interference of fine-grained background noise and gathering similar information of coarse-grained context works.
SpecAugment only considers the introduction of different types of noise to diverse the training samples to expect a more robust model. Mixup directly mixes different samples, which may produce some samples that do not match the reality. 
In the proposed EDC, information fusion in the event-related local area  preserves the event-related features while eliminating the interference caused by too much background noise.

\begin{table}[b]\footnotesize 
	\setlength{\abovecaptionskip}{0.0cm}   
	\setlength{\belowcaptionskip}{-0.45cm}   
	\renewcommand\tabcolsep{2.0pt} 
	\centering
	\caption{Performance of different methods on the same model.}
	\begin{tabular}
	{
	p{0.4cm}<{\centering} |
	p{2.8cm}<{\centering} | 
	p{0.9cm}<{\centering}
	p{0.9cm}<{\centering}|
	p{0.9cm}<{\centering}
	p{0.9cm}<{\centering}
	} 
\hline
		\multirow{2}{*}{\#} 
		& \multirow{2}{*}{\textit{Conditioning method}} 
		& \multicolumn{2}{c|}{\textit{AUC (\%) of OM}}
		& \multicolumn{2}{c}{\textit{AUC (\%) of AM}}  
		\\
		
		\cline{3-6}
		
		 &  & \textit{DCASE} & \textit{CHiME} & \textit{DCASE} & \textit{CHiME} \\ 
		  
		  \hline
		  
		 1 &  \textit{None} & 90.55 & 91.28 & 90.71 & 91.42 \\
		  
		 2 &  \textit{Self-attention} \cite{transformer} & 84.36 & 91.91 & 84.89 & 92.16 \\

		 3 &  \textit{SpecAugment} \cite{spec} & 92.61 & 94.83 & 93.06 & 94.94 \\
		  
		 4 &  Mixup \cite{mixup} & 93.44  & 95.76 & 94.67 & 95.56 \\
		   
		 5 &  \textit{Feature weighting} \cite{feature_w} &  94.22 & 96.07 & 94.27 & 96.08 \\
		  
		 6 &  \textit{Proposed EDC} & \textit{\textbf{95.20}} & \textit{\textbf{96.93}} & \textit{\textbf{96.03}} & \textit{\textbf{98.10}} \\

\hline
	\end{tabular}
	\label{tab:models}
\end{table}

\noindent
\textbf{• RQ3}: How about the details of the model trained with EDC?

Figure~\ref{figure-f} shows the model trained with EDC successfully recognizes the classes of acoustic events, and detects the probability of events in each frame. The event probability curve predicted by the model matches acoustic representations, even the onset and offset of each sub-event are detected well.  
The reason may be the enhancement of local information based on the event-related adaptive time window in EDC plays the role of clustering, so results in Figure~\ref{figure-f} show the feature map after EDC will gather locally similar information well.
However, the key information related to events is smoothed out by the self-attention (Figure~\ref{figure-f} (b)), which makes it difficult for the model to distinguish which are the core representations of events and which are irrelevant background.

\label{ssec:figure1}
\begin{figure}[t] 
	\setlength{\abovecaptionskip}{0.05cm}    
	\setlength{\belowcaptionskip}{-0.6cm}   
	\centerline{\includegraphics[width = 0.5  \textwidth]{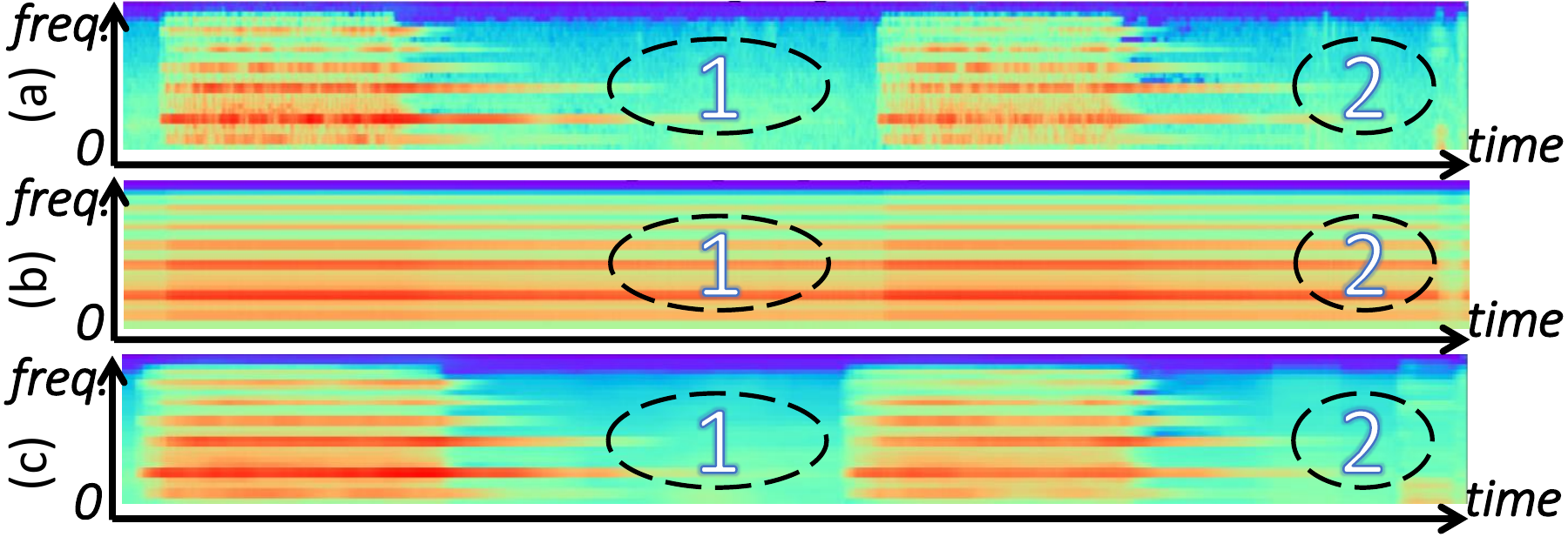}}
	\caption{Subgraph: (a) Log mel features; (b) Bottleneck features from the model with self-attention; (c) Features after EDC. }
	\label{figure1}
\end{figure}

To gain a more intuitive insight into the differences between the feature tuned by EDC and the feature learned by the trainable self-attention, Figure 5 shows an example from the same audio clip.
Compared with the input log mel features, in Figure~\ref{figure1}, representations learned by the self-attention overly enhance parts 1 and 2. 
For these two intervals, information that did not originally belong to them was added, and boundaries between acoustic events representations and background were smoothed out.
Unlike self-attention, EDC extracts information from neighboring frames related to the event, instead of from global scope.
Figure~\ref{figure1} (c) shows that after EDC, the information in the event-related local part is enhanced and boundaries between events and backgrounds are clearly preserved.

\section{CONCLUSION}
\label{sec:CONCLUSION}
This paper proposes EDC, which is a deterministic feature transformation to gather event-related local information and enhance boundaries between events and backgrounds.
The core of EDC is to adaptively select the frame-related attention range based on acoustic features to focus on the event-related local information.
The experiments on two real-life polyphonic audio datasets of a total of 17 classes of events show that EDC performs well when compared with SpecAugment, Mixup, learnable feature weighting and self-attention in the AEC task. 
Future work will further analyze the impact of similarity probability distribution $P_{f_{\omega i}}$ and attenuation coefficient $D(n)$ in the proposed EDC on diverse acoustic events in real life.

\vfill\pagebreak

\bibliographystyle{IEEEtran}

\bibliography{mybib} 

\end{document}